\documentclass[12pt]{article}
\usepackage{graphicx}

\begin{document}

\def\onehalf{{\textstyle \frac12}}
\def\ii{{\rm i}}
\def\dd{{\rm d}}
\def\ssr#1{{\scriptscriptstyle\rm #1}}
\def\of#1{{{\scriptstyle(}#1{\scriptstyle)}}}
\def\jour#1#2#3#4{{\it #1{}} {\bf #2}, #3 (#4)}
\def\lab#1{\label{eq:#1}}
\def\rf#1{(\ref{eq:#1})}
\def\Lie#1{\hbox{\sf #1}}
\def\rectangulo#1{\centerline{\framebox{\LARGE #1}}}
\def\ket#1{\,\vert{#1}\rangle}
\def\bra#1{\langle{#1}\vert}
\def\pmbf#1{\rlap{$#1$}{}\hskip-0.5pt{#1}}
\def\abs#1{{\scriptstyle|}#1{\scriptstyle|}}
\def\aabs#1{{\scriptscriptstyle|}#1{\scriptscriptstyle|}}
\def\tsty#1#2{{\textstyle\frac{#1}{#2}}}
\def\of#1{{{\scriptstyle(}#1{\scriptstyle)}}}
\def\oof#1{{{\scriptscriptstyle(}#1{\scriptscriptstyle)}}}
\def\ssty#1{{{\scriptscriptstyle(#1)}}}
\def\brof#1{{{\scriptstyle[}#1{\scriptstyle]}}}

\newcommand{\be}{\begin{equation}}
\newcommand{\ee}{\end{equation}}
\newcommand{\bea}{\begin{eqnarray}}
\newcommand{\eea}{\end{eqnarray}}

\begin{center}
{\LARGE Discrete Bessel functions and transform} \\[20pt] 	
 {Kenan Uriostegui}\footnote{Posgrado en Ciencias F\'{\i}sicas, 
Universidad Nacional Aut\'onoma de M\'exico} and {Kurt Bernardo Wolf}\\[15pt]
{Instituto de Ciencias F\'{\i}sicas\\
Universidad Nacional Aut\'onoma de M\'exico\\Av.\ Universidad s/n, 
Cuernavaca, Morelos 62251, M\'exico}
\end{center} 

\vskip25pt

\begin{abstract}
We present a straightforward discretization
of the Bessel functions $J_n(x)$ to 
discrete counterparts ${B}_n^\ssty{N}(x_m)$, 
of $N$ integer orders $n$ on $N$ integer points $x_m\equiv m$,
that we call {\it discrete\/} Bessel functions. 
These are built from a Bessel integral generating function, 
restricting the Fourier transform over the circle to $N$ points. 
We show that the discrete Bessel functions satisfy several
linear and quadratic relations, particularly Graf's 
product-displacement formulas, that are exact analogues
of well-known relations between the continuous 
functions.
It is noteworthy that these discrete Bessel functions
approximate very closely the values of the continuous
functions in ranges $n+\abs{m}<N$. For fixed $N$, this 
provides an $N$-point transform between functions of 
order and of position, $f_n$ and $\widetilde{f}_m$, 
which is efficient for the Fourier analysis 
of finite decaying signals.
\end{abstract}

%-----------------------------------------

\section{Introduction: Discrete Bessel functions\label{sec:one}}  

Fourier-Bessel analysis originates from the radial part of 
two-dimensional Fourier analysis. Solutions of the wave
equation in cylindrical coordinates yield the functions 
known collectively as cylinder functions. Of these, we shall
be particularly interested in the Bessel functions of the
first kind and of integer order, defined from a plane-wave 
decomposition that provides their generating function
\cite[KU120(13)]{GR},
\be
	e^{\ii x\sin\varphi} = J_0(x) + 2\sum_{k=1}^\infty
		J_{2k}(x)\cos(2k\varphi) +2\ii\sum_{k=0}^\infty
		J_{2k+1}(x)\sin((2k{+}1)\varphi),
				\lab{genfJ}
\ee				
for $x\in{\cal R}$ real and $\varphi\in{\cal S}^1$ the circle.
Seen as Fourier sine and cosine series, this provides an 
expression for the coefficient functions, as
\be
	J_k(x)=\frac1{2\pi}\int_{-\pi}^\pi \dd\varphi\,
		\exp(\ii x \sin\varphi)
		\Big[C_k\cos(k\varphi)-\ii S_k \sin(k\varphi)\Big],
			\lab{Besssc}
\ee
where we used
\be 
	C_k:=\vert\cos(\onehalf k\pi)\vert=\left\{ 
		{{\,1,\  k \hbox{ even},}
		\atop 0,\  k \hbox{ odd},}\right.\quad
	S_k:=\vert\sin(\onehalf k\pi)\vert=\left\{ 
		{{0,\  k \hbox{ even},}
		\atop 1,\  k \hbox{ odd}.}\right.
		\lab{CS}
\ee

In Fourier analysis, a well-known strategy to discretize
the Fourier integral transform over a circle to an $N$-point 
cyclic finite Fourier transform, is to replace integrals by finite sums 
over $N=2j+1$ equidistant points $\varphi_k$ on the 
circle, through
\bea 
	\frac1{2\pi}\int_{-\pi}^\pi \dd\varphi\,f(\varphi)
		&\longrightarrow& \frac1{2j{+}1}\sum_{k=-j}^j
		f(\varphi_k),\lab{disc1}\\
	\varphi_k:=\frac{2\pi\,k}{2j{+}1} &\hbox{so}& 
		\Delta\varphi=\varphi_{k+1}-\varphi_k
				=\frac{2\pi}{2j{+}1}. \lab{disc2}
\eea
We consider $j$ to be integer, and thus $N$ odd.

In Ref.\ \cite{Biagetti-etal} the authors
proposed to discretize the Bessel function 
from its integral definition \rf{genfJ} to the 
$N$-point sum; however, their results
are incomplete for not having respected the 
difference between the even and odd orders, with
values over different sets of points over the circle. 
We thus propose here to expand the definition of 
{\it discrete Bessel functions\/} as
\bea
	B_n^\ssty{N}(x_m)&:=&
		\frac1{2j{+}1}\sum_{k=-j}^j
		\exp(\ii x_m \sin\varphi_k)
				\Big[C_n\cos(n\varphi_k)-\ii S_n \sin(n\varphi_k)\Big],
			\lab{Besscii} \\	
		&=&\frac1{2j{+}1}\sum_{k=-j}^j
		\exp(\ii m \sin\varphi_k)\times\left\{ 
		{{\phantom{-\ii}\cos n\varphi_k,\  n \hbox{ even},}
		\atop -\ii\sin n\varphi_k,\  n \hbox{ odd},}\right.
		 	\lab{defBessDisc}		 	   
\eea
where $n$ and $m\equiv x_m$ are integers; their range, 
initially the set of all integers, can be reduced to 
$n\in\{0,1,\ldots,\,N{-}1=2j\}$, or to 
$m\in\{-j,-j{+}1,\ldots,\,j\}$, due to 
the symmetries
\be
	\begin{array}{c}
	 B_n^\ssty{N}(m)=
	(-1)^n B_{-n}^\ssty{N}(m) = (-1)^n B_n^\ssty{N}(-m) \quad
	\hbox{real},\\[3pt]  
	\! B_n^\ssty{N}(0)=\delta_{n,0}. \end{array}	\lab{symmet}
\ee

In Sect.\ \ref{sec:two} we show that beyond
superficial similarities, the discrete Bessel functions 
$B_n^\ssty{N}(m)$ exhibit various other properties that 
are exact counterparts of those satisfied by the continuous Bessel 
functions $J_n(x)$. This includes linear relations that
are proven straightforwardly, and the quadratic relation
known as Graf's formula \cite{Graf-origin} in its various 
forms and special cases.
 
The feature that initially caught attention is shown
in Sect.\ \ref{sec:three}, where we compare the actual numerical 
values of the discrete and continuous Bessel functions.
Although it is clear from the beginning that a properly
written limit $\lim_{N\to\infty}B_n^\ssty{N}(m)$ 
should return the continuous Bessel function $J_n(x)$, 
the approximation provided by the discrete Bessel 
function is surprisingly close in a region of
the integer grid of indices $\of{n,m}$. 
The differences between the two for $0\le n+m<j 
\approx\onehalf N$ are of the order of 
$\Delta_{n}^{\!\!\oof{80}}< 10^{-16}$ for $N=161$; they are smaller 
when farther from the upper edge of that region.

We shall call the $N\times N$-matrix ${\bf B}^\ssty{N}=
\Vert{B}_n^\ssty{N}(x_m)\Vert$ in \rf{Besscii}--\rf{defBessDisc} 
simply as the {\it discrete\/} Bessel transform kernel.
Note that among Bessel functions at $x=0$, only $J_0(0)=1\neq0$;
the discrete Bessel matrix also has at its apex ${B}_{0}^\ssty{N}(0)=1$,
with the rest of its column ${B}_{n>0}^\ssty{N}(0)=0$.
This discrete Bessel matrix 
is nonsingular, so in Sect.\ \ref{sec:four} we can define a 
{\it discrete Bessel transform\/} between an $N$-point function
$f_m\equiv f(x_m)$ and its partial Bessel coefficients $\widetilde{f}_n$. 
The $N\to\infty$ limit of this transform is not granted, however.

The quest for discrete analogues of the cylinder functions
has both computational and analytical interest, and has
been approached in several ways, from an early definition
in Ref.\ \cite{RHBoyer}, to recent work based on 
difference equations postulated as analogues to the Bessel
differential equation \cite{Bohner-Cuchta,Slavik}. The
resulting definitions are not equivalent to that of
Ref.\ \cite{Biagetti-etal} nor the one we study here.
Therefore we emphasize in the concluding Sect.\ \ref{sec:five}
that, beyond the many analytic properties and computational
applications that cylinder functions have enjoyed,
the present discrete analogue and its associated
transform may have not yet been regarded.

%--------------

\section{Linear and Graf discrete Bessel identities}
							\label{sec:two}

The discrete Bessel functions $B_n^\ssty{2j+1}(m)$ 
in \rf{Besscii} obey analogues of several well-known
identities satisfied by the continuous Bessel functions 
$J_n(x)$. The Fourier series transform over the circle can be
straightforwardly discretized to the transform over 
$2j+1$ points in linear expressions.  For brevity, 
we shall henceforth omit the upper index, understanding
that $N=2j+1$; formulas will also 
simplify upon the introduction of the so-called
\be
	\hbox{Neumann factor: }\ 
		\varepsilon_n := 2-\delta_{n,0}=\left\{ \begin{array}{rl}
		1, & n=0, \\ 2, & n\neq 0. \end{array}\right. \lab{Neumaneps}
\ee
It is then straightforward to write and prove the sum of 
even orders of the discrete Bessel functions, as
\be 
	\sum_{n=-j}^{j} B_{2n}(m) 
	= \sum_{n=0}^{j} \varepsilon_n \, B_{2n}(m)
	= B_0(m) + 2\sum_{n=1}^j B_{2n}(m)=1,
				\lab{lin1}
\ee 
which can be compared with the corresponding equation
in \cite[Eq.\ WA44, p.\ 934]{GR}. This is a particular
case (for $\varphi_k=0$) of the linear discrete Bessel 
summations for odd and even orders,
\bea
	\sum_{n=0}^j \varepsilon_n B_{2n}(m)\cos(2n\varphi_k)	
		&{=}& \cos(m\sin\varphi_k),  \lab{Bcos}\\
	\sum_{n=0}^j  B_{2n+1}(m)\sin((2n{+}1)\varphi_k)	
		&{=}& \onehalf \sin(m\sin\varphi_k),   \lab{Bsin}
\eea
which has been proven for the four cases of $k$ and $n$ 
even or odd, using trigonometric sum identities. They
can be compared with \cite{GR}, Eqs.\ KU120(14) and (15), on 
p.\ 935, respectively. 

Regarding quadratic expressions, a sum found by Neumann
in 1867 for integer orders, was extended by Graf in 1893 
to all real orders, and known since as Graf's formula 
\cite[Sec.\ 7.6.2, Eq.\ (6)]{Graf-origin}. Its
group-theoretic origin is the linear transformation 
of spherical harmonics $Y_{\ell,m}(\theta,\phi)$ by
Wigner-$d$ functions under rotations around the $y$-axis,
contracted for $\ell\to\infty$ \cite{Graf}. In that 
limit both spherical harmonics and Wigner $d$-functions 
become Bessel functions and particularly yield
\be
	\sum_{n=-\infty}^\infty 
			J_{n}(x)\, J_{n'-n}(x')=J_{n'}(x+x'),  \lab{Grafeq}
\ee
which can be seen as a displacement and convolution
of arguments and indices.

The discrete Bessel functions $B_n(m)$ 
in \rf{Besscii}--\rf{defBessDisc} satisfy a corresponding 
formula for $N=2j{+}1$, that is
\be
	 \sum_{n=-2j}^{2j} 
			B_{n}(m)\, B_{n'-n}(m')
			=B_{n'}(m+m').  \lab{B-Grafeq}
\ee
To prove this relation, we directly replace the discrete
functions from \rf{Besscii}, keeping in mind the parity and
periodicity properties \rf{symmet}, which allow the sum to
be over the $N$ terms, as $\sum_{-j}^j$ or $\sum_0^{2j}$. 
The left-hand side of this discrete Graf formula is
\bea
	\sum_{n=-2j}^{2j}\!\!B_{n}(m)\, B_{n'-n}(m')
	\!\!\!&=&\!\!\!\frac1{(2j{+}1)^2} \sum_{n=-2j}^{2j}\!\! \sum_{k,k'=-j}^j
		\exp\ii( m \sin\varphi_k + {m'}\sin\varphi_{k'})\nonumber\\
    &&{}\times \Big[C_n\cos(n\varphi_k)-\ii S_n \sin(n\varphi_k)\Big],
    		\lab{Graf2}\\
    &\times& \Big[C_{n'-n}\cos((n'{-}n)\varphi_{k'})
    	-\ii S_{n'-n} \sin ((n'{-}n )\varphi_{k'})\Big]. \nonumber
\eea
The sum over $n$ shifts over to the two factors that house
the parities in \rf{CS}, which then separate into two cases,
for $n'$ even or odd, to reconstruct the right-hand side of
the discrete Graf formula. 

For $n'$ even, the sum over $n$ becomes, 
\be 
	\begin{array}{l} \displaystyle
	\sum_{n=-2j}^{2j}\Big[C_n C_{n'-n}\cos(n\varphi_k) \cos((n'{-}n)\varphi_{k'})
		+ S_n S_{n'-n} \sin(n\varphi_k)\sin ((n'{-}n )\varphi_{k'})\Big]\\
		\displaystyle{}\qquad\qquad{}= (2j{+}1)\cos(n'\varphi_k)\,\delta_{k,\,k'},
				\end{array}  \lab{nppar}
\ee
while for $n'$ odd the sum yields
\be 
	\begin{array}{l} \displaystyle
	-\ii\!\!\!\sum_{n=-2j}^{2j}\!\!\!\Big[C_n S_{n'-n}  \cos(n\varphi_k)\sin ((n'\!{-}n )\varphi_{k'})
	+ S_n C_{n'-n}\sin(n\varphi_k) \cos((n'\!{-}n)\varphi_{k'})\Big]\\
		\displaystyle{}\qquad\qquad{}= 
		-\ii\,(2j{+}1)\sin(n'\varphi_k)\delta_{k,\,k'}.
		 				\end{array}  \lab{npnon}
\ee
This brings the left-hand side of \rf{B-Grafeq} closer to that 
of $B_{n'}(m{+}m')$,
\be
	\frac1{2j{+}1} \sum_{k,k'=-j}^j
		\exp\ii( m \sin\varphi_k + {m'}\sin\varphi_{k'})
     \Big[\widetilde C_{n'}^{k,k'}\cos(n'\varphi_k)
    		-\ii\widetilde S_{n'}^{k,k'} \sin(n'\varphi_k)\Big],
    		\lab{Graf2b}
\ee
where \rf{nppar} and \rf{npnon} contribute to the coefficients
\be 
	\widetilde C^{k,k'}_{n'}= \left\{  \begin{array}{ll}
		 0, & n' \hbox{ odd}, \\ 
		 \delta_{k,\,k'}, & n' \hbox{ even,} \end{array} \right.
		 \qquad 
	\widetilde S^{k,k'}_{n'}= \left\{  \begin{array}{ll}
		 \delta_{k,\,k'}, & n' \hbox{ odd},\\ 
		  0, & n' \hbox{ even.} \end{array} \right.
   \lab{contldes}
\ee
These factors placed in the double sum \rf{Graf2}
reduce terms to those with $k=k'$ into a single sum, where the
two exponents join to render $(m+m')\sin{\varphi_k}$, and the
left-hand side of \rf{B-Grafeq} has become indeed
\be 
	\frac1{2j{+}1} \sum_{k=-j}^{j}
		\exp(\ii\, \of{m{+}m'} \sin\varphi_k) 
     \Big[C_{n'}\cos(n'\varphi_k)-
     	\ii S_{n'} \sin(n'\varphi_k)\Big]
     	=B_{n'}(m{+}m').
				\lab{fiDem}
\ee				

The symmetries \rf{symmet} indicate that the $4j+1$
summands in $\sum_{n=-2j}^{2j}$ can be reduced to 
$N=2j+1$ summands by introducing the Neumann factor
$\varepsilon_n$ in \rf{Neumaneps},
Thus, for $n'=0$ and $m=-m'$, Eq.\ \rf{B-Grafeq} can
be written in the forms
\be 
	\sum_{n=-2j}^{2j} \Big[B_n(m)\Big]^2
		= \sum_{n=0}^{2j}\varepsilon_n \Big[B_n(m)\Big]^2
	=	\Big[B_0(m)\Big]^2 + 
		2\sum_1^{2j} \Big[B_n(m)\Big]^2
	=1,   \lab{sumcuad}
\ee
that correspond to well-known formulas for Bessel 
functions with infinite sums. 
The basic Graf formula that is valid for the discrete
Bessel functions yields several of its versions
under the symmetries listed in \rf{symmet}. 
We should mention that all formulas in this
Section have also been verified numerically.

%---------------

\section{The discrete-to-continuous approximations}
				\label{sec:three}
				
\begin{figure}[t]
\centering  
\centering{\includegraphics[width=1.0\linewidth]{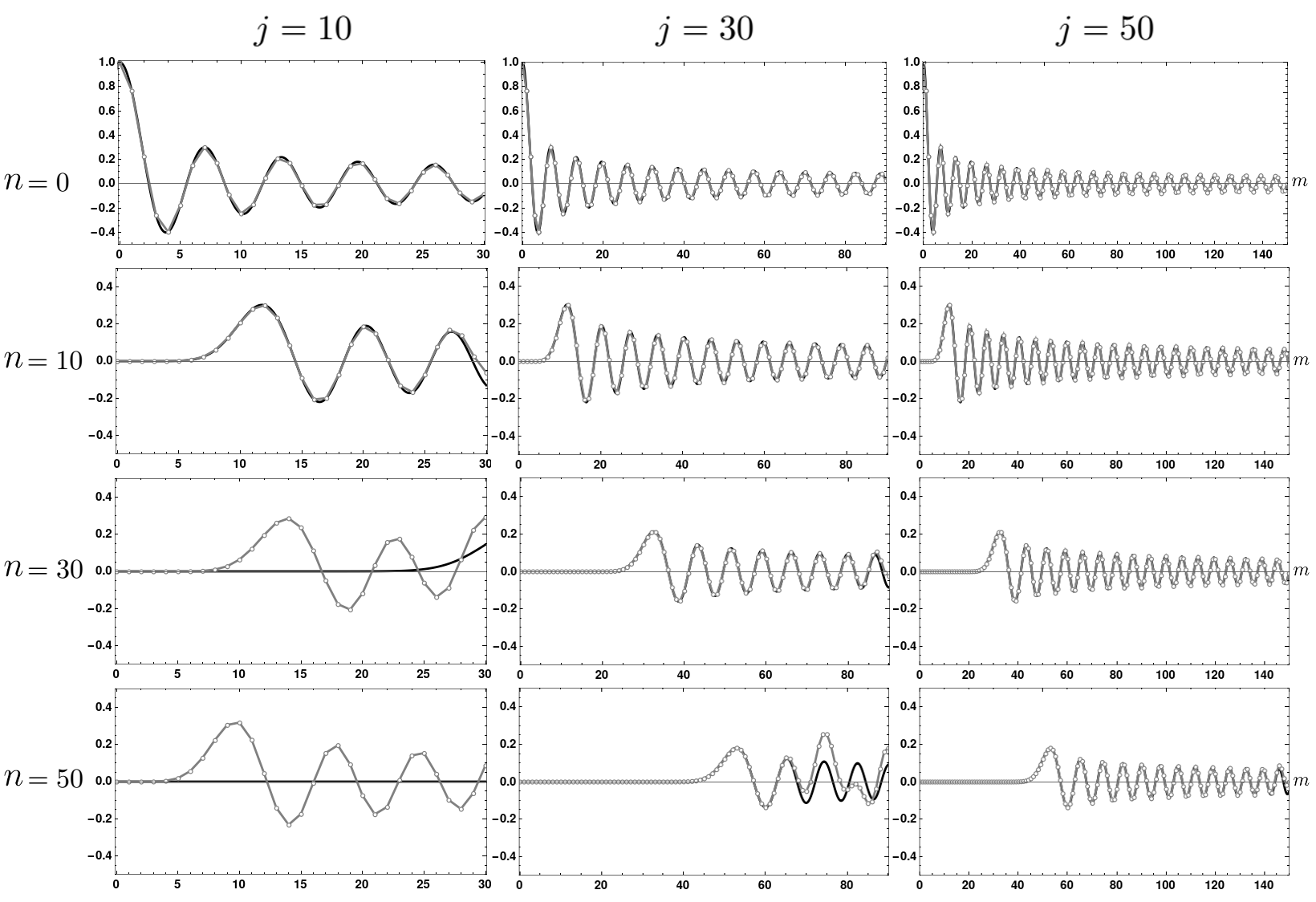}}
%\rectangulo{${\displaystyle j\in\{10,30,50\},\atop n\in\{0,10,30,50\},\quad m\in[0,4j]}$}
\caption{Comparison of values of the discrete 
Bessel functions $B_{n}^\ssty{2j+1}\!(m)$
(open circles) and the continuous Bessel
functions $J_n(m)$ (lines) in the same ranges.
We show the cases for $j\in\{10,\,30,\,50\}$, (i.e., 
$N\in\{21,\,61,\,101\}$ discrete points), for Bessel 
orders $n\in\{0,\,10,\,30,\,50\}$, over the ranges
$m\in[0,\,4j\,{=}\,2N-2]$ of their argument.
The continuous lines are gray where the difference 
between the discrete interpolation and the continuous 
Bessel values is less than $10^{-16}$ and replaced 
by heavy black lines where it is greater.} 
\label{fig:aproxms}
\end{figure}

Figure \ref{fig:aproxms} shows the values of the discrete 
and the continuous Bessel functions for various values 
of $j,\,n$ and $m$. In this section we 
indicate the approximate equalities by
$B_n(m)\widetilde{=} J_n(m)$ and report estimates
for ranges in the integer grid $(n,m)$ adjacent
to the origin. For $N=2j{+}1$ differences, we 
measure the merit of the approximation through 
the mean quadratic error,
\be 
	\Delta_{n}^{\!\ssty{N}}:=\frac1N\sum_{m=0}^{N-1}
		\Big( J_n(m)-B_n^\ssty{N}(m)\Big)^2.  
		\lab{error}
\ee

The space of a discrete system of $N=2j+1$ points 
$\{x_m\}_{m=0}^{2j}$ is spanned by a basis set of 
$N$ independent functions $J_n(x_m)$. The graphs 
in Fig.\ \ref{fig:aproxms} contain intervals of $m$ 
beyond $[0,2j]$, and up to $4j=2N-2$. Good matches
between discrete and continuous Bessel values 
in the grid $(n,m)$ are seen to lie in the first 
quadrant for $n+m\le2j$; in the interval $[0,2j]$,
the mean square error between values is 
$\Delta_{n}^{\!\ssty{321}}<10^{-16}$, for $j=160$.
In the next Section we shall use this feature to
define a discrete Bessel transform between functions
of position $m$ and mode $n$ for any given integer $j$.

\begin{figure}[t]
\centering  
\centerline{\includegraphics[width=1.0\linewidth]{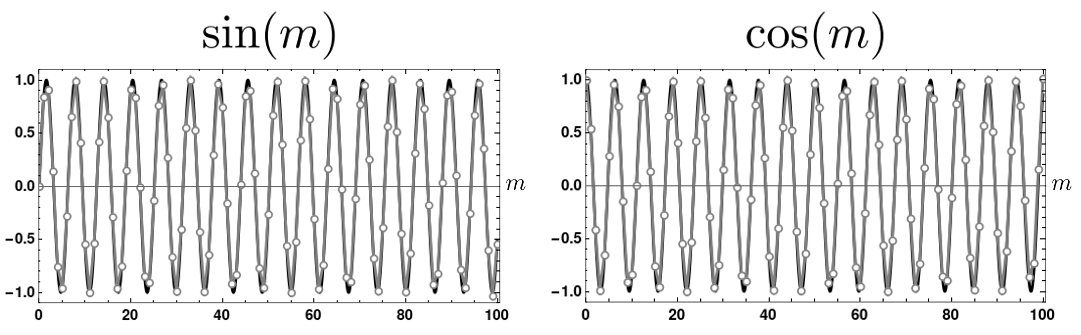}}
%\rectangulo{$\sin(m) \qquad \cos(m)$}
\caption[]{Comparison of the left-hand sides of Eqs.\ 
\rf{BSin} and \rf{BCoss} for integer $m\in[0,100]$ 
(open circles), and $\sin(m)$ and $\cos(m)$ for continuous $m$
(line) in the same range; here, $j=50$ and $N=101$.}  
\label{fig:sin-cos}
\end{figure}

When we enlist other well known formulas that are
valid for continuous Bessel functions, and replace
them by their discrete version we also find matches
with similar approximations. Among them we find
\bea 
	2\!\sum_{n=0}^j (-1)^n B_{2n+1}(m) 
		&\widetilde{=}& \sin (m),  \lab{BSin}\\
	\sum_{n=0}^j\varepsilon_n (-1)^n B_{2n}(m) 
		&\widetilde{=}& \cos (m),  \lab{BCoss}
\eea
that can be compared with Eqs.\ \cite[WH(1,2), p.\ 934]{GR}. 
We point to the 
fact that in \rf{BSin}--\rf{BCoss}, the argument of
sine and cosine is {\it integer\/} 
$m\in\{-j,-j{+}1,\ldots,j\}$. To compare
these functions of discrete $m$ with the 
continuous functions of $m$, we show both
in Fig.\ \ref{fig:sin-cos}. The mean square
errors there are of the order $\approx 10^{-6}$.

\begin{figure}[t]
\centering  
\centerline{\includegraphics[width=1.0\linewidth]{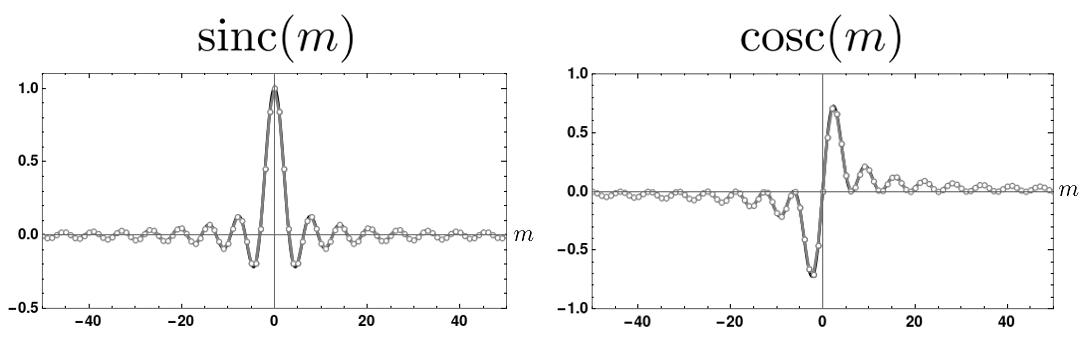}}
%\rectangulo{sinc \quad cosc}
\caption[]{Comparison of values of the discrete 
and continuous Bessel functions, left- and right-hands 
of Eqs.\ \rf{Bsssin} and \rf{Bcccos}, as before,
with open circles and continuous lines,  in the same 
range for $j=50$, $N=101$.} 
\label{fig:sinc-cosc}
\end{figure}

For other expressions in Fig.\ 
\ref{fig:sinc-cosc} we show, for $n=1$ and 
$m\in[-j,j]$, the approximations
\bea
	\sum_{k=0}^j  B_{1}(m\cos\varphi_k)\cos\varphi_k	
		&\widetilde{=}& \frac{\sin m}{m}=:\hbox{sinc}\,m,   \lab{Bsssin}\\
	\sum_{k=0}^j  B_{1}(m\cos\varphi_k)	
		&\widetilde{=}& \frac{1-\cos m}{m} =:\hbox{cosc}\,m.  \lab{Bcccos}
\eea
that also hold with a mean square error less than $10^{-6}$. 

%----------------------------

\section{Discrete Bessel transform and inverse} 
				\label{sec:four}
				
In Section \ref{sec:one} we introduced the $N\times N$ 
{\it discrete Bessel matrix\/} ${\bf B}=\Vert B_{n,m}\Vert$,
$B_{n,m}:= B_n(m)$. Any finite set of $N$ linearly independent 
vectors can be used to define an $N$-dimensional vector space;
although we cannot prove linear independence here, numerical 
verifications of $\det{\bf B}\neq0$ support this very
plausible conclusion. 
Hence, given a function of $N$ positions $f_m$, the matrix 
$\bf B$ will transform this into a function 
$\widetilde{f}_n$ of $N$ {\it modes}. The inverse matrix 
${\bf C}:={\bf B}^{-1}$ then recuperates the original 
function of positions,
\be 
	\widetilde{f}_n:=\sum_{m=0}^{N-1} B_{n,m}\,f_m,
	\quad  f_m=\sum_{n=0}^{N-1} C_{m,n}\widetilde{f}_n,
	\qquad {\bf CB}={\bf1}.
		\lab{dir-inv}
\ee

The elements of the matrix $\bf B$ closely approximate
the values of $J_n(m)$ on the integer grid region $0\le n+m 
\le N-1$. We saw that $B_{n,0}=\delta_{n,0}=J_n(0)$, 
and  we know that also for $0\le m\le n{-}1$, both the
continuous $J_n(m)$ and the discrete Bessel 
functions are very small. They start oscillating with 
a maximal amplitude just beyond $m\approx n$, and 
decrease as $\sim m^{-1/2}$ along the position $m$-axis. 
The matrix $\bf B$ is thus effectively upper-triangular 
and the value of its determinant will be approximately 
given by the product of its $N$ diagonal elements,
$D_N:=\det{\bf B} \approx \prod_{n=0}^{N-1} J_n(n)$.
This determinant quickly becomes very small: for 
$N=\{5,\,11,\,21,\,51\}$, one finds $D_N<\{10^{-2},\,
2{\times}10^{-6},\,7{\times}10^{-14},\,10^{-39}\}$.
Although this does not negate the existence of the
discrete Bessel transform \rf{dir-inv}, it effectively 
will reduce the numerical stability of computations
for larger $N$'s, and will preclude the existence of an 
$N\to\infty$ limit.
				
We should remark here that series or integral transforms
with kernels involving Bessel functions, are quite
distinct from \rf{dir-inv}. The Hankel-$n$ transform kernel
is $\sim (pq)^{1/2}J_n(pq)$, between function spaces
$f(q)$ and $\widetilde f(p)$; the two-dimensional 
Fourier-Bessel series with the drum harmonics has the kernel 
$\sim J_n(k_{n,m}r)\,e^{\ii n\phi}$
in polar coordinates, with integer $n,m$ and 
$k_{n,m}$ being frequencies allowed	by circular 
boundary conditions. Further series detailed in Watson's 
treatise  \cite{Watson} are the Neumann-$c$ series with 
kernel $\sim J_{n+c}(r)$, the Kapteyn-$c$ series with 
$\sim J_{n+c}(\of{n{+}c}r)$, and the Schl\"omlich-$\mu$
series with $J_n(\mu r)$, to transform between functions
$f_n$ with $n$ integer and $\widetilde f(r)$ on
continuous $r$. As can be seen, none has the simple
structure of the discrete, $N\times N$ Bessel matrix in 
\rf{dir-inv}.
				
%----------------------------

\section{Concluding remarks}   \label{sec:five}

After trigonometric functions, cylinder ---and in 
particular Bessel--- functions can be seen as
the most important in mathematical physics.
These Bessel functions also have important
group-theoretic properties as irreducible 
representation matrix elements for the
Euclidean groups of rotations and translations
\cite{Talman}.

The trigonometric functions are the basis for
three distinct transforms: the Fourier integral
transforms, Fourier series, and the finite Fourier
transforms. The three relate through continuous
limits and discretizations of the real line $\cal R$, 
the integers $\cal Z$, and on ${\cal Z}_N$, the 
integers modulo $N$. Such a discretization
of the generating functions was used here for $J_n(m)$
on $\cal Z$ to define $B_n(m)$ on ${\cal Z}_N$. In
fact, the position $m$ need not be integer; as
occurs in the finite Fourier case, the $N$ phases
on the circle can be shifted freely. The values
of the discrete Bessel functions in the figures
can be similarly defined on an $m$-line in $[0,N{-}1]$
or extended beyond.
On the other hand, the order $n$ cannot but be integer,
because for non-integer $n$, $\vert J_n(0)\vert$ is
infinite.

The interest on the discrete Bessel functions \rf{defBessDisc}
as proposed in Ref.\ \cite{Biagetti-etal} is that the discretization
of decaying, or radial wave propagation, or scattering in two 
or more dimensions, could be profitably seen as a superposition 
of Bessel normal modes, whose starting location and decay give physical
meaning to those modes. Cylinder functions other than the Bessel
functions of the first kind will be investigated elsewhere.

%----------------------------------

\section*{Acknowledgments}

We thank the support of the Universidad Nacional Aut\'onoma
de M\'exico through the PAPIIT-DGAPA project AG--100119
{\it \'Optica Matem\'atica}.

\end{document}